# Dry Late Accretion inferred from Venus' coupled atmosphere and internal evolution

C. Gillmann[1], G. J. Golabek[2], S. N. Raymond[3], M. Schönbächler[4], P. J. Tackley[5], V. Dehant[6], V. Debaille[1]

[1]Laboratoire G-Time, Université Libre de Bruxelles, Brussels, Belgium.

[2]Bayerisches Geoinstitut, University of Bayreuth, Bayreuth, Germany.

[3]Laboratoire d'Astrophysique de Bordeaux, CNRS and Université de Bordeaux, Pessac, France

[4]Institute of Geochemistry and Petrology, ETH Zürich, Zürich, Switzerland.

[5]Institute of Geophysics, ETH Zurich, Zürich, Switzerland.

[6]Royal Observatory of Belgium, Brussels, Belgium.

## Abstract

**The composition of meteoritic material delivered to the terrestrial planets after the end of core formation as late accretion remains contentious. Because the evolution of Venus' atmospheric composition is likely to be less intricate than the Earth's, we test implications of wet and dry late accretion compositions, using present-day Venus atmosphere measurements. Here we investigate the long-term evolution of Venus using self-consistent numerical models of global thermochemical mantle convection coupled with both an atmospheric evolution model and a late accretion N-body delivery model. Atmospheric escape is only able to remove a limited amount of water over the history of the planet. We show that late accretion of wet material exceeds this sink. $CO_2$ and $N_2$ contributions serve as additional constraints. A preferentially dry composition of the late accretion impactors is in agreement with observational data on $H_2O$, $CO_2$ and $N_2$ in Venus' present-day atmosphere. Our study suggests that the late accreted material delivered to Venus was mostly dry enstatite chondrite, conforming to isotopic data available for Earth. Our preferred scenario indicates late accretion on Venus contained less than 2.5% wet carbonaceous chondrites. In this scenario, the majority of Venus' and Earth's water has been delivered during the main accretion phase.**

## Main Text

The volatile inventory of terrestrial planets is of prime importance because it not only affects the atmosphere, the surface conditions and the potential habitability, but also influences mantle dynamics, the tectonic regime and volcanic outgassing[1,2]. Uncertainties remain mostly due to limited data on the early volatile reservoirs[3], and uncertain delivery scenarios[3,4,5].

After core formation ceased, late meteoritic material delivery occurred on Earth, contributing as late accretion (LA) an additional ≈0.5-1.5% of Earth's mass[5,6,7]. While advances have been made to constrain Earth's LA, the exact nature of that late material remains debated when it comes to other bodies. It must account for the highly siderophile elements (HSE) abundances in Earth's mantle, but might also have contributed to water delivery to Earth[8]. While volatile-rich bodies like carbonaceous chondrites (CC) have been considered as an ideal candidate, especially if the Earth's mantle was left dry after the Moon forming giant impact[8], recent geochemical results argue for LA on Earth that was mainly composed of dry material[9] like enstatite chondrites (EC)[10].

Numerical results show that LA also occurred on Venus[11], but in absence of samples from Venus, isotopic studies are limited to the planet's atmosphere[1,2,12]. Venus' past is poorly constrained, however, it has been suggested that Venus experienced a more straightforward evolution of its volatile inventory[1,13]. Such an evolution would involve less volatile recycling, as Venus shows no sign of a biosphere, standing water or global plate tectonics.

While liquid water has been suggested to be stable on Venus, under specific conditions, for long periods of time[14], without a magma ocean, there is no sufficiently strong oxygen sink to remove the oxygen resulting from the eventual evaporation of a large water ocean caused by a runaway greenhouse effect[1,15,16]. Conversely, conditions during the global magma ocean phase are not compatible with the condensation of large bodies of water at the surface[1,17]: condensation of water means the end of the magma ocean and the end of the strongest oxygen sink. Thus, large bodies of liquid water are difficult to accommodate on Venus both early and late during its evolution. It is therefore likely that Venus has been desiccated since its early evolution[18], the extent and timing of which are still poorly constrained, and recent surface conditions do not allow for water condensation and associated enhanced weathering[1,17].

Venus' massive $CO_2$ atmosphere also contains about as much $CO_2$ as all atmospheric, crustal and mantle reservoirs on Earth. Thus the present-day atmosphere of Venus is closer to its primitive inventory of $CO_2$, while on Earth, a comparable amount has been incorporated into carbonates and the mantle[1]. Without volatile recycling, Venus is also less likely to develop Earth-like plate tectonics[19], which would explain the lack of a magnetic dynamo[20].

We propose a quantitative self-consistent model of Venus, focussing on the evolution of its volatile inventory since late accretion. We have combined N-body LA calculations[11] with a state-of-the-art coupled planet interior and atmospheric evolution model[21]. We model the consequences of LA on Venus' volatile history and exchanges of three species ($H_2O$, $CO_2$ and $N_2$) between the planetary interior and atmosphere.

**The early evolution of Venus**

The evolution of Venus starts with its accretion, when the bulk of the mass of the planet is delivered[1]. Both accretionary impacts and atmosphere result in temperatures high enough to create a magma ocean[22]. Due to hydrodynamic escape the primordial atmosphere thins over a timescale of a few tens of million years[12,15]. For a slow to moderate solar rotation period modelling work on isotopic fractionation of noble gases indicates the loss of a terrestrial water ocean every $10^5$-$10^7$ years. Models also suggest that the magma ocean solidifies rapidly over a period of $\approx 1$ Myr[17], when atmospheric water content becomes small[17,18] due to hydrodynamic escape. The mantle enters the phase of solid-state convection (Fig. 1) that we model here. Afterwards, late accretion delivers the final part of the planetary mass via impacts[5,6].

Late accretion impacts have three main effects (all considered here): delivery of volatiles to the planet[8], impact heating[23] and erosion of the atmosphere[24]. We use N-body simulations to study late accretion, with various simulated growth histories that are consistent with the masses of present-day terrestrial planets. Since it is possible that the size and number of impactors have consequences for the evolution, we have tested four distinct LA impact scenarios (Fig. 2). Those represent median cases based on several hundred N-body simulations[11]: a single very large impact (Scenario A; $R$ = 1819 km)[25], 9 large impacts (Scenario B; $R >$ 500 km), 82 medium-sized collisions (Scenario C; $R >$ 125 km), and finally 244 small impactors (Scenario D; $R >$ 50 km). We note however that, given the top-heavy nature of the impacting size distributions, the mass tends to be dominated by the few largest bodies. The timing of LA on Venus is still debated, thus onset times ranging from 50 to 150 million years (Myr)[26] after the formation of Ca-Al-rich inclusions (CAIs) are used, marking the start of the evolution models. Impactors have either an EC-like composition[27,28] (0.1% $H_2O$, 0.4% $CO_2$, 0.02% $N_2$) or a CC composition[29,30] (8% $H_2O$, 4% $CO_2$, 0.2% $N_2$), thus determining the amount of volatiles delivered by LA to the atmosphere of Venus[31]. Intermediate ordinary chondrites (OC) compositions are also considered (see supplementary information).

**The consequences of volatile delivery**

During LA volatile delivery, strong escape mechanisms limit the accumulation of volatiles in the atmosphere. Thermal (hydrodynamic) escape dominates[12] the early evolution. As atmospheric water is photo-dissociated, early hydrodynamic escape removes $H_2$ efficiently during the first few 100 Myr. Since removal of oxygen[1,12,32,33] by hydrodynamic escape is inefficient, it accumulates in the early atmosphere. However, the present-day atmosphere of Venus displays very little water or oxygen. Therefore, it must have been lost during planetary evolution[1]. During the main accretion stage, oxidation of the molten magma ocean can trap oxygen efficiently[1,15,34], but from the LA stage onwards, this sink only exists in the form of oxidation of the newly emplaced lava flows[35], thus its consequences on oxygen removal are limited (see Supplementary information).

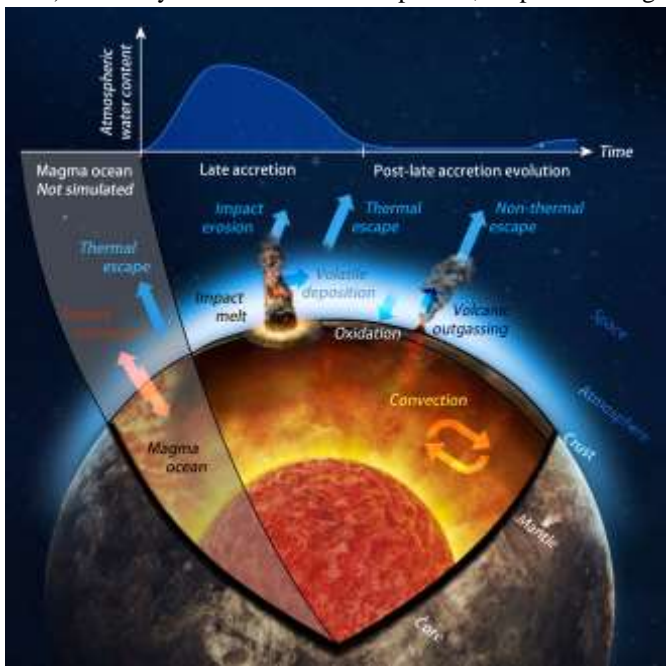

**Figure 1**: **Volatile exchanges on Venus**. Mechanisms affecting the water content of Venus' atmosphere during the long-term evolution of the planet.

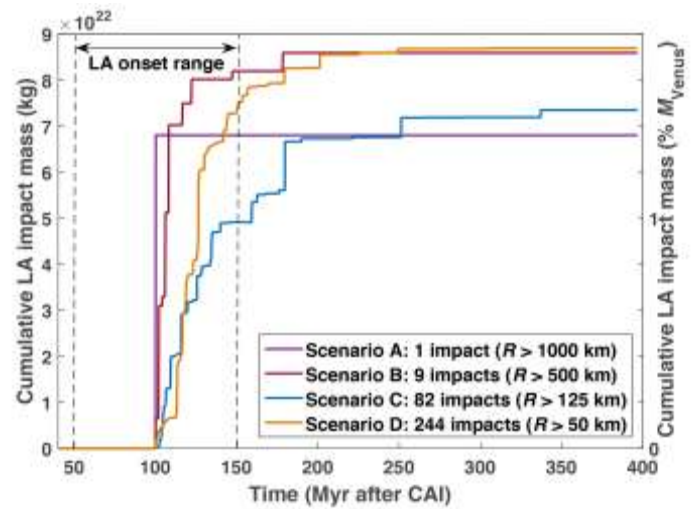

**Figure 2: Late accretion scenarios**. Four distinct LA impact scenarios are used in the coupled interior-atmosphere models assuming a LA beginning at 100 Myr after the formation of CAIs. The considered variation in the LA onset time is shown by the dashed line box. The total mass of Venus LA is found to be larger than Earth's by a factor 2-3.

Non-thermal escape removes less mass than hydrodynamic escape. However it removes oxygen and affects the long-term evolution of the atmosphere. The balance between oxygen left in the atmosphere by LA, hydrodynamic escape and the long-term sink of non-thermal escape (plus putative oxidation mechanisms[34]), defines the upper limit of LA volatile delivery and its composition.

Since the removal of oxygen occurs on the timescale of billions of years, the entire evolution of Venus starting with LA needs consideration. To this end, the LA impact scenarios are incorporated into coupled atmosphere-interior evolution models that follow the model layout by Gillmann et al.[21] considering volcanic outgassing as additional volatile source (see Methods). We model the evolution of the atmospheric volatile inventory starting after the end of the magma ocean stage until present-day. Comparison between the final state of the simulated atmosphere and the present-day atmosphere composition enables us to constrain the volatile content of the LA material.

Our main constraint is the bulk composition of the present-day atmosphere of Venus, which is well established[1], with 11 x $10^{18}$ kg $N_2$ (2 bar[36]), $10^{16}$ kg $H_2O$ (2 x $10^{-3}$ bar[34,37]) and 4.69 x $10^{20}$ kg $CO_2$ (90.6 bar[38]) (see Methods). Other major constraints available to narrow down Venus evolution scenarios are the noble gas isotopic ratios and the D/H ratio. Noble gas isotopic ratios ($^{36}Ar/^{38}Ar$ and $^{20}Ne/^{22}Ne$) are thought to indicate early thermal escape has fractionated those species[1,12]. However this constraint on escape is weak, since it has been shown that no single solution meets the $^{36}Ar/^{38}Ar$ and $^{20}Ne/^{22}Ne$ fractionation[12]. We still use this constraint to assess hydrodynamic escape fluxes. The upcoming Venera D mission may improve noble gases isotopic measurements and provide tighter constraints on Venus evolution. The D/H atmospheric ratio shows an enrichment of deuterium relative to Earth of a factor 100 indicating water loss from the Venus atmosphere[39]. It is governed by long-term non-thermal escape and its interaction with volcanic degassing sources of unknown composition[40], with additional effects from hydrodynamic escape[41] and impacts[42]. Moreover, measurements show that the D/H ratio in the atmosphere of Venus changes with altitude[39], limiting its value as a constraint. Thus, modelling of D/H ratio evolution is not feasible here.

**Estimating maximum late accretion delivery**

At the start of the evolution model, after the magma ocean has solidified, the initial water content of Venus' atmosphere is assumed to have been completely removed during earlier evolution[1]. Since our aim is to estimate the maximum volatile content of LA, we chose ideal initial conditions with no pre-existing water. Any oxygen remaining at the onset of the simulation would also need to be extracted from the atmosphere by the limited existing volatile sinks. This would imply a reduced LA volatile delivery and reinforce our results. An early $CO_2$ atmosphere exists due to magma ocean outgassing, while $N_2$ is mostly trapped at the hot surface[43]. Initial volatile pressures at the start of the LA are chosen to be 0 bar $N_2$ [43] and $H_2O$, and 65 bar $CO_2$ [15]. $CO_2$ degassing of the magma ocean is considered to be efficient due to its low solubility in magma, and a thick $CO_2$ atmosphere is formed early on[1,17]. The LA stage is characterized by volatile delivery due to impacts and subsequent volatile loss. Water delivery is dominated by impactors with CC composition (if present), because of their high water content, while $N_2$ and $CO_2$ are deposited in large amounts by both CC and EC. Impact erosion effects[44] are noticeable but insufficient to prevent volatile build-up. Volcanic outgassing, including that caused by impact melting, also remains a second-order effect. The late evolution of atmospheric water content (from the end of LA to present-day) is dominated by atmospheric escape and volcanic degassing, also causing the slow accumulation of $N_2$ and $CO_2$ in Venus' atmosphere.

The major factor that limits $H_2O$ evolution models is the "escape envelope", corresponding to the maximum amount of oxygen that can be removed by atmospheric escape between a given time and present-day[45] (Fig. 3a). It is obtained by cumulating losses based only on non-thermal escape processes from present-day, back to any given time in the past[46]. Thus models where the delivered water and consequent $H_2O$ equivalent pressures exceed the escape envelope are not able to lose sufficient amounts of volatiles to meet present-day observations and will be rejected. All models are evaluated against the present-day volatile content of the atmosphere. As an additional constraint, Venus must lose all molecular oxygen that it accumulated during its early evolution due to hydrodynamic escape[1,46,47].

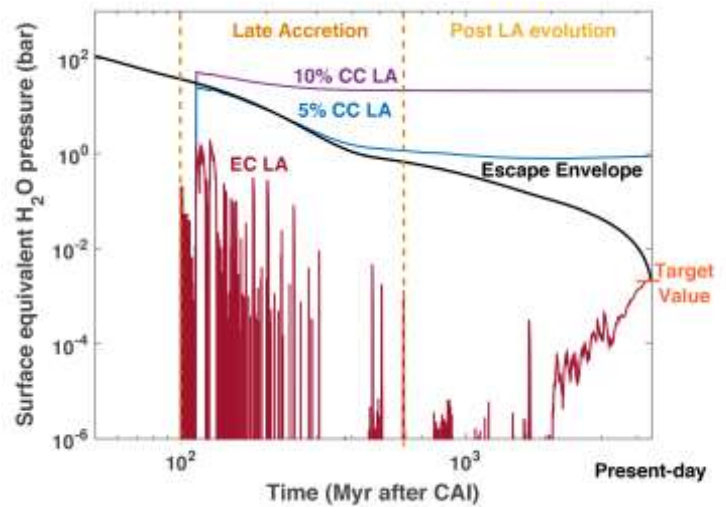

**Figure 3: Evolution of water in the atmosphere of Venus.** Time evolution of $H_2O$ in the Venus atmosphere for MAX conditions assuming different LA compositions, labelled as CC material percentage of the total LA mass delivery. LA scenario D starting at 100 Myr after CAI formation is used.

The

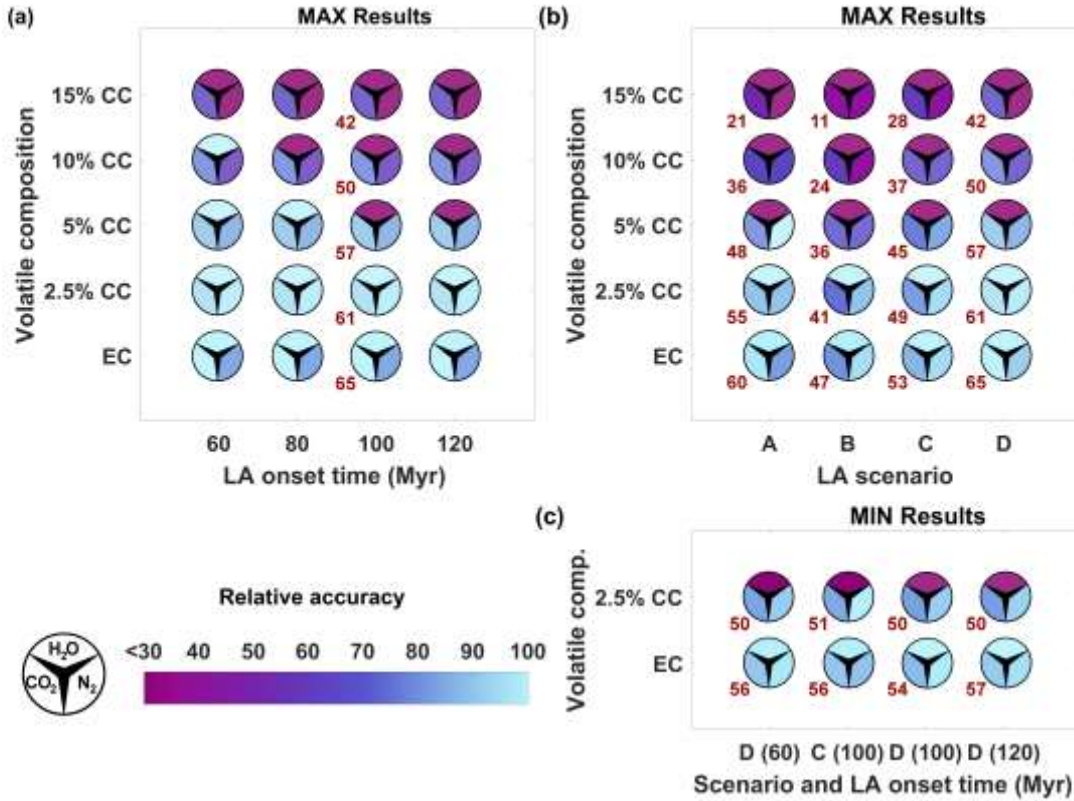

**Figure 4: Consequences of late accretion volatile content on present-day Venus atmosphere.** Agreement (relative accuracy) between model results and the present-day volatile content of Venus' atmosphere. (a, b) assuming strong atmospheric escape (MAX parameter set), (c) using weak atmospheric escape (MIN parameter set). (a) Scenario D starting at 60-120 Myr after CAIs and different LA compositions ranging from an EC end-member to 15% of CC material. (b) LA scenarios A-D with LA onset at 100 Myr assuming the same variation of LA compositions as in (a). (c) Selection of simulations employing scenarios C and D with different LA onset times. A minus sign indicates a negative difference. Red numbers indicate the post magma ocean $CO_2$ pressure (bar) required as initial condition to increase the $CO_2$ agreement at present-day to 100%.

evolution of water abundances shown in Figure 3 represents an equivalent water concentration based on the fluxes of O atoms from both water delivery and losses. For $H_2O$, two distinct cases arise: either (i) $H_2O$ is removed, the atmosphere desiccates and only limited volcanic degassing during late evolution replenishes it, or (ii) it remains in the atmosphere resulting in higher equivalent water pressures (>1 bar) for the rest of the evolution, not fitting present-day observation of Venus. For $CO_2$ and $N_2$ the progression from one scenario to the other is much smoother (Extended data figure 1).

The choice of non-thermal escape parameters is therefore crucial. Three main sets were tested, controlling the magnitude of volatile loss. The parameter set that allows the largest potential volatile deposition from LA (i.e. the most favourable to wet LA), therefore being considered as the upper limit for LA volatile content, is denoted MAX. This end-member is characterized by maximum atmospheric escape[46], minimal outgassing[21], and the use of the tangent plane model for impact-induced atmospheric erosion[44]. The parameter set called MIN causes the smallest loss and therefore the driest LA and is considered to be the most realistic[1]. A third scenario with intermediate volatile loss is called MED. Extended data Figure 2 and Figure 3 show the same evolution with MED and MIN parameters.

Figures 4a and 4b show the relative agreement between model results and observations, using the relative accuracy of simulations ($\alpha = 1 - |M_{S,i} - M_{obs,i}|/M_{obs,i}$, where $i$ denotes the volatile species considered, while $M_s$ and $M_{obs}$ are the simulated and observed masses, respectively). Figure 4 summarizes model results based on different LA onset times (Fig. 4a) and the four LA impact scenarios (Fig. 4b) for the cases allowing for a maximum volatile concentration in LA bodies. For comparison, Fig. 4c shows results for selected models using more realistic water losses. Based on $H_2O$ only, calculations show that a maximum of 2.5% of wet LA material consistently satisfies present-day conditions for a LA beginning around 60-120 Myr (Fig. 4a). This would be the equivalent of a single CC impactor with ≈500 km radius. The onset time of LA greatly affects the results in MAX cases, due to the additional losses it allows for. Late LA onsets (>150 Myr) are unlikely to fit the atmosphere composition data. Assuming an earlier LA onset (while keeping the same

LA mass) the results for water imply a viable amount of up to 10%-15% of wet LA material. On the other hand, $N_2$ and $CO_2$ evolution provide additional constraints, since their loss rates are lower than those of water[1,46]. Correspondingly, early volatile-rich LA models that satisfy the $H_2O$ content are unable to simultaneously meet present-day $N_2$ and $CO_2$ measurements. In the case of LA containing 10% CC material, the present-day amount of $CO_2$ is 22% off (≈20 bars), while $N_2$ is 40% off (≈0.7 bar). Those discrepancies are larger than the total contribution of volcanic outgassing for $CO_2$ and $N_2$, respectively (Extended data figure 4). Therefore, such a LA composition is in disagreement with the measured atmospheric composition data, and the excess of volatiles cannot be attributed to excessive outgassing. Therefore, the maximum contribution of CC in LA is <2.5% for LA onset times >100 Myr and <10% for earlier LA onsets. Thus, our results do not allow us to constrain the LA onset time on Venus, but rather, indirectly, the impact time of the last major wet impactor[45].

Figure 4b illustrates the interplay of total LA mass and the number of impactors. Larger masses result in more volatile delivery. On the other hand, for a given mass, many small impacts cause more atmospheric erosion than a smaller number of larger impactors. Thus, Figure 4b implies that the size-frequency distribution of LA impactors has only minor effects on our conclusions, especially for water. In particular, we do not expect our results to change if the LA impactors had a size distribution that was less top-heavy and closer to the present-day asteroid size distribution. Figure 4c shows water loss estimates based on the MIN parameter, representing more accurately our current understanding of solar EUV and water loss evolution[1]. MIN models are extremely restrictive and have a tighter limit on water delivery. The escape envelope shows a small slope, thus not able to constrain the LA onset time very well. This parameter set requires a dry LA scenario, with less than 2.5% CC (and >97.5% EC material), regardless of LA onset time and implies that impactors may have been significantly desiccated[48].

In conclusion, the best fit between models and present-day Venus atmosphere shows that LA on Venus was primarily composed of EC-like (dry) meteorites with at most 5% of the total mass exhibiting a CC (wet) composition. Venus and its atmosphere have therefore not received any major volatile delivery after the end of the magma ocean. Numerical models suggest that LA could have been homogeneous in the early solar system with comparable compositions, including that of water[11,13,49]. In that case, our conclusions agree with previous suggestions based on isotopic studies for both Earth and Mars material[9,10,50] and indicate that the majority of volatiles was delivered to the terrestrial planets during planet formation, and thus, in the case of Earth, before the Moon-forming giant impact.

**Methods.**

The coupled atmosphere-interior model considers four main aspects defining the long-term evolution of Venus: (i) volcanic degassing of volatiles, (ii) atmosphere escape sinks, (iii) surface conditions evolution, and (iv) late accretion (LA) impacts. We test three distinct parameters sets: MAX parameters (see main text; allowing for maximum LA volatile contribution), MED parameters (supplementary material, average parameters) and MIN parameters (see main text; low volatile loss, allowing for minimum LA volatile contribution). See Extended Data Tables 1 and 2. The list of models is given in Supplemental information Table 1. Based on present-day knowledge the MIN parameter set is considered to be the most realistic[5]. The MAX parameter set leads to extreme escape cases; it is used for the numerical models discussed in the main text because the extreme nature of this parameter set also ensures that results are robust to parameter variations. All atmospheric volatile pressures are expressed assuming that this specific gas is the only one in the atmosphere, and are therefore a proxy of mass. As a reference, one terrestrial ocean corresponds to 265 bar, $1.39 \times 10^{21}$ kg and a global equivalent layer of 2900 m depth.

**Numerical modelling of thermochemical convection.** Mantle dynamics is modelled for 4.5 Gyr, from the end of the magma ocean stage until present-day. We use the version of the code StagYY[51] described in Armann and Tackley (2012)[52], Gillmann and Tackley (2014)[53] and Gillmann et al. (2016)[21]. The tectonic regime of the planet is not prescribed, but it is calculated self-consistently. In the present simulations Venus never exhibits Earth-like plate tectonics. Annexe 1 of Supplementary Information contains a full description.

**Melt eruption and volatile outgassing.** Melt generated during mantle convection and reaching the surface of Venus contributes to volatile outgassing of $N_2$, $CO_2$ and $H_2O$. The full description of the outgassing process and parameters can be found in Supplementary Information Annexe 1. Extended data Figure 4 shows impact and volcanic contributions.

**Solar Evolution.** The long-term evolution of extreme UV (EUV) flux from the Sun and solar wind affect atmospheric escape[15,46,54]. EUV flux decreases with time according to a power law[55], assuming that the early Sun was a slow to moderate rotator, as suggested by recent studies[1,56]. Luminosity increases with time, according to the Faint young Sun hypothesis[57,58] from 70% of present-day luminosity to the present-day value.

**Hydrodynamic/thermal escape.** It describes the response of a primitive, hydrogen-rich atmosphere subjected to high intensity EUV radiation. The atmosphere expands, hydrogen escapes and heavier species can be dragged along only as long as the driving H flux persists. We use the energy-limited model of hydrogen escape described in Gillmann et al. (2009)[47] and based on previous work[34,59,60,61,62,63]. Maximum escape fluxes calculated for hydrogen are $\approx 3 \times 10^{31}$ s$^{-1}$, consistent with estimates from recent work[15,46]. Photodissociation splits water molecules into H and O atoms.

As H escape is much more efficient than O escape, hydrodynamic escape generates an accumulation of O (possibly as much as 85 bars O[62,63]) that needs to be removed from the atmosphere: early on only, it can be stored in the magma ocean. At the onset of the model, all primordial water is thus assumed to have been removed from the atmosphere[1]. Any oxygen remaining in the atmosphere reduces potential water delivery during the LA. Additionally, this means most of the thermal escape of O and $CO_2$ occurs during the magma ocean phase, before the onset of our model and therefore controls the initial volatile inventory. $CO_2$ losses have been suggested to be low (<10 bar[15,46,64,65]). $N_2$ was protected due to the cooling effect of $CO_2$ in the upper atmosphere[1]. Later on, during and after LA, oxygen and $CO_2$ are lost primarily by non-thermal escape mechanisms[1,46,47], due to the lack of a late long-lived H-rich atmosphere[1]. Therefore, due to its very limited losses during and after LA, $CO_2$ thermal escape rates are not considered in the model. The error between modelled and observed present-day $CO_2$ pressures is determined and a corrected initial pressure necessary to reduce the error to zero is calculated (see Fig. 4).

**Non-thermal escape.** It covers the interaction of upper atmosphere particles with the high-energy radiation from the Sun like for example the EUV flux. Mechanisms involved are photochemical reactions[66], sputtering[67], ion pick-up[68] and plasma instabilities[54]. Non-thermal escape dominates late evolution (3.5 Ga-present) but also plays a role earlier on. We reconstruct past non-thermal escape fluxes in two stages: Before and after 3 Ga, as detailed in the next two paragraphs. The cumulated effect of oxygen loss over time is shown as the escape envelope in Figure 3. Non-thermal $CO_2$ losses are very small over the history of Venus[21,46].

**Recent water loss.** It is calculated from O escape rates. Late evolution rates for the last 3 Gyr are calculated as in Gillmann et al. (2014, 2016)[21,53], based on simulations and present-day measurements[64] of atmospheric escape rates at solar minimum and solar maximum EUV conditions. An interpolation between those two values indicates how escape varies over a solar cycle depending on the EUV flux. Solar maximum conditions of a present-day solar cycle correspond to the solar minimum EUV conditions 2.5-2.8 Gyr ago[55]. Thus, we have access to the variation of escape rates for minimum solar conditions over the last 2.8 Gyr, which we extrapolate to mean escape rates used for the time period 0-3 Gyr before present-day. We use a cumulated present-day escape rate of 6 x $20^{25}$ s$^{-1}$, (combining all O loss mechanisms) close to the maximum limit derived for present-day conditions[4]. The reconstruction is consistent with simulations from Kulikov et al. (2006)[46].

**Late accretion-era O excess removal.** Early non-thermal escape is calculated using numerical simulations[46] of ion-pick-up loss flux based on a particle model[69] and taking into account the stronger early solar activity (EUV about 100 times the present-day value 4.5 Gyr ago[58]). Specifically, we use case 2b (moderate solar wind activity). This estimate corresponds to the effect of ion-pick-up only. The total escape, accounting for all mechanisms is several times higher[64]. MAX (main text and figures) and MED (extended data) parameter cases use a multiplication factor of 5 and 2.5, respectively (case 2b in Kulikov et al., 2006[46]), to account for higher and lower escape ranges. Low escape (case 4 in Kulikov et al., 2006[46]), representing recent advances in the characterization of solar evolution[1,70,71], is used in MIN parameter cases. Further models use the lowest possible multiplicative factor (6) able to accommodate LA scenario D with enstatite chondrite composition. Extended data Table 2 compares parameters for these three cases. While thermal escape remains unaffected by the presence of a magnetic field, non-thermal escape is affected. It is currently uncertain whether Venus had a magnetosphere in the past[72], and there is currently no sign of a present-day magnetosphere. It has been suggested that, if Venus exhibited Earth-like conditions, it would probably still maintain a magnetic field at present-day[72]. On the other hand, if Venus had a stratified core, it would never have generated a magnetic field[73]. Given present knowledge of the effects of magnetic fields, it is unknown whether the presence of a magnetosphere would result in decreased or increased atmospheric losses[74]. In case the atmosphere is shielded, our conclusions are reinforced. Assuming an end-member scenario where Venus had a magnetosphere during its entire evolution resulting in maximum additional $H_2O$ losses[74] leads to an increased maximum contribution of CC material in our models of up to 15%, 10% and 3% assuming MAX, MED and MIN parameter sets, respectively. However $N_2$ and $CO_2$ limits remain unchanged (Figure 4).

**Surface oxidation sink.**

Oxidation of the surface of Venus has been proposed as a possible oxygen sink[34,41] and would be consistent with the suggestion[75] of an oxidized basaltic surface. We have performed additional numerical calculations that include surface oxidation to assess its consequences on the results of the model. We consider the following chemical reaction[76]: 2 $FeSiO_3$ + ½ $O_2$ -> $Fe_2O_3$ + 2 $SiO_2$

Freshly erupted lava is able to react with oxygen while it is still hot. However since the diffusion of oxygen in hot lava is slow, the process is inefficient. Results of the simulations and full description of the calculations are available in Annexe 1 of the supplementary information. The effect of oxidation is noticeable, but only a small amount of up to a few bar of oxygen at most is removed this way from the atmosphere.

**Surface conditions.** Surface temperature is calculated using a one dimensional, vertical, radiative-convective grey atmosphere model as described in Gillmann et al. (2014)[53]. The atmosphere is considered to be in hydrostatic equilibrium and in a regime similar to present-day. The equilibrium temperature of the planet evolves with solar luminosity. $CO_2$ and $H_2O$ are the only greenhouse gases considered here. Over time, the $CO_2$ content of the atmosphere only varies moderately compared to present-day conditions. Water atmospheric concentration is more variable, but within the bounds of previously tested models[21,53]. Surface temperatures calculated were found to be consistent with canonical values[1]. Changes in surface temperature affect convection[53], however consequences on the volatile repartition are marginal, due to the low outgassing efficiency. Simulations with a constant surface temperature show 3% of change in atmospheric volatile inventory.

**Late accretion scenarios.** We use a series of N-body simulations from Raymond et al.[11] to generate a population of impacting bodies for Venus' late accretion. These simulations provide plausible evolutionary histories that match our current interpretation of the constraints. In our simulations, Jupiter and Saturn were included on orbits consistent with their gas-driven orbital migration[77,78,79]. The terrestrial "leftovers" were included on orbits consistent with simulations of terrestrial planets formation[80,81]: the orbital semi-major axis was randomly chosen between 0.5 and 1.7 au, the eccentricity between 0.1 and 0.6, and the inclination between zero and 20 degrees. The mass of the leftovers were drawn from size distribution consistent with Bottke et al.[82] and parameterized as $dN/dD \sim D^{-q}$, where $N$ is the number of objects of a given diameter $D$, and $q$ was chosen to be 1, 1.5 or 2. For the main set of simulations[11], terrestrial leftovers (minimum diameter of 1000 km) represent a population totalling 5% of Earth's mass, which was found to deliver the correct amount of late accretion mass to Earth[11]. Scenarios A and B adopted the stochastic late accretion framework of Bottke et al.[82], who suggested that a top-heavy mass distribution could explain why Earth experienced a more substantial late accretion than that inferred for the Moon or Mars[7,83]. Due to the limited number of impactors in scenarios A and B small number statistics has to be considered. For this purpose, these two scenarios employ median cases based on hundreds of N-body simulations from Raymond et al.[11]. We also consider LA scenarios involving smaller impactors[84]. Therefore, we include two additional, higher-resolution simulations (scenarios C and D) for this work with minimum diameters of 100 and 250 km, covering a wide range of size-frequency distributions for the LA. All scenarios used here are found to deliver the correct amount of LA material to the Earth (0.25-0.75% of Earth's Mass)[5,6,7], satisfying HSE constraints. As a rule, they deliver a slightly larger mass as late accretion to Venus. This is consistent with findings by Jacobson et al.[26]. Our simulations start with fully formed terrestrial planets and a population of leftovers, implying that "time zero" for these simulations was the last giant impact on any of the terrestrial planets, (the Moon-forming impact, roughly 50-150 Myr after CAIs for Earth[85,86]). It has been shown that an early last giant impact on Venus implies a larger total LA mass of up to $\approx 20\%$ of the planet's mass[26,73]. Such LA scenarios result in additional volatile delivery and are unlikely to accommodate CC material. We performed numerical calculations using a modified scenario D where LA onset is at 20 Myr and all impactor masses are scaled up, so the total LA mass is 10% of the Venus mass (see table 3). Results indicate no more than 10% CC material is permitted, to match present-day $H_2O$ atmospheric content, while $N_2$ and $CO_2$ deliveries never match with present-day observations. We find the differences between models A to D to be small (Fig. 4), and mainly due to differences in total mass. This leads us to expect that the size distribution of the LA impactors is not a critical factor for atmospheric volatile content and that our model would still match the constraints for a different distribution more characteristic of the present-day asteroid belt.

**Late accretion composition.** Two compositional end-members are chosen: Enstatite chondrites (EC)[27,28] or carbonaceous chondrites (CC)[29,30,87]. Reducing the water content of the enstatite chondrite end-member does not affect the maximum mass of CC compatible with atmospheric constraints by more than 1%, because water delivery is dominated by the CC bodies. An EC $H_2O$ concentration reduced by a factor 10 has been tested without notable change to the results (Table 3). The effects of variations in CC water content are shown in Supplementary Information Table 1; our conclusions remain unaffected.

We performed several calculations with an ordinary chondrite (OC) end-member instead of a CC composition[88,89], using 1% $H_2O$, 1% $CO_2$, 0.01% $N_2$. Lower water concentrations make this reservoir very similar to the EC end-member and resulting evolutions are close to the EC-only cases featured in Table 3. Results for OC LA scenarios indicate a correspondingly higher fraction of the wet end-member with less than 45% OC with MAX parameter set, 30% with MED and 10% for MIN parameter set. This is (especially for MED) in line with results obtained by Brasser et al. (2017)[90].

In scenarios B-D, compositional variation is handled by assigning part of the LA impactors, up to the desired mass fraction a CC composition, while the bulk of impactors have an EC composition. Those impactors with CC composition are always assumed to collide with Venus during the first 20 Myr of the specific LA scenario to maximize volatile loss. For control models and scenario A, all impactors have a uniform intermediate composition. Control models (Supplementary Information Table 1) exhibit marginal changes, indicating they can accommodate 1-3% (weight) more CC material.

**Impact-induced atmosphere erosion.** Impacts can cause atmospheric loss due to atmospheric entry, ejecta material and vapour plume generation. Here, we use two approaches to model impact erosion[21]: (i) the tangent plane model[91,92] and (ii) results from SOVA hydrocode simulations[24,93,94]. The tangent plane model predicts the removal of all atmospheric content above a plane tangent to the surface of the planet at the impact location, thus each collision removes $\approx 10^{-3}$ times the mass of the atmosphere of Venus. We consider for the MAX and MED parameter models that all LA impactors are large enough to cause atmospheric escape as described by the tangent plane model. SOVA hydrocode results imply lower erosion rates and are used for MIN parameter models. For large bodies ($R > 100$ km) SOVA-derived erosion efficiencies are reduced by a factor $\approx 4$ compared to the tangent plane model. Here, the total amount of atmospheric erosion by a certain LA scenario is primarily dictated by the number of impacts. We have neglected ground-motion atmospheric erosion[95], which may be significant for large impacts (scenario A) but negligible for smaller ones (scenarios B-D). This mechanism would increase impact erosion in scenario A, making it comparable to those of scenarios C-D, thus leading to more uniform results across all mass-size distributions.

**Impact delivery of volatiles.** Volatile delivery depends on impactor composition (see above) and the fraction of the volatiles released during the collision that remain in the atmosphere afterwards. This deposition efficiency must take into account the portion of impactor material retained and outgassing of volatiles from the impactor material. Numerical simulations[24,93,96] suggest that a minor fraction (10%) of the projectile can be lost back to space as ejecta material. We consider only vaporized impactor material to contribute to outgassing of volatiles. We use a conservative value for the vaporized fraction of 40%[96,97,98], leading to a deposition efficiency of 36%. In case a larger part of the projectile is vaporized, the conclusions presented in the main text are reinforced. We tested low-end deposition efficiency factors (12%; see table 3) with similar conclusions leading to a maximum EC fraction of ≈10% for MAX parameters (less than 5% for MED and 2.5% for MIN).

**Impact heating.** Shock heating generates a high temperature region in the planetary interior. Mantle volume and temperature increase in this thermal anomaly is treated as in Gillmann et al. (2016)[21,23], taking into account impact velocities higher than the escape velocity[99,100]. Only impactors large enough ($R > 150$ km) to penetrate the lithosphere[21,23] are considered to have a significant effect on the temperature structure of the planet's convecting interior. Smaller impactors only contribute to the atmosphere's evolution. Crustal vaporization is neglected. Compared to 3D geometry the 2D geometry introduces an overestimation of the relative mantle volume affected by thermal anomalies[21]. In 2D and 3D, the ratio of affected volume to mantle volume varies with $R^2_{anomaly}/(R^2_{Venus}-R^2_{core})$ and $R^3_{anomaly}/(R^3_{Venus}-R^3_{core})$, where $R_{Venus}$, $R_{core}$, and $R_{anomaly}$ are the radii of the planet, its core and the isobaric core of the thermal anomaly, respectively[101]. The ratio between these two values approximates the overestimation caused by geometry. For large impacts ($R > 500$ km) it is ≈3, while for medium-sized collisions (150 km ≤ $R$ ≤ 500 km) the factor is ≈7. Therefore, only one in three large impactors and one in seven medium-sized ones are considered to create a thermal anomaly.

**Data availability.** The data that support the findings reported in this article are available as follows: Code outputs of N-body simulations (impactors and collisions parameters) are available from figshare, with the identifier DOI: 10.6084/m9.figshare.11829621. Data generated for the models displayed in figures (equivalent pressure evolutions) are available from figshare, with the identifier DOI: 10.6084/m9.figshare.11829621. Datasets generated during the current study as present-day Venus atmosphere composition for the full complement of models is available in supplementary material. **Code availability.** The convection code StagYY is the property of PJT and Eidgenössische Technische Hochschule (ETH) Zürich. It is available on request from PJT (paul.tackley@erdw.ethz.ch). The N-body model MERCURY, used for LA scenarios, is available at https://github.com/4xxi/mercury.

*Corresponding author:* All queries should be sent to the corresponding author (cgillmann@ulb.ac.be)

*Acknowledgements.* We thank Fabio Crameri for providing the perceptually-uniform colour map name used in Fig.4 (Crameri, F. Geodynamic diagnostics, scientific visualisation and StagLab 3.0, Geosci. Model Dev., **11**, 2541-2562, 2018,). We thank Dave Rubie for his comments. We also thank Ramon Brasser, Kevin Zahnle and an anonymous reviewer for their constructive comments. CG, VDeh and VDeb were supported by BELSPO PlanetTOPERS IUAP programme and ET-HoME Excellence of Science programme. VDeb thanks the FRS-FNRS and ERC StG ISoSyC FP7/336718. MS acknowledges the National Center for Competence in Research "PlanetS" supported by the Swiss National Science Foundation (SNSF). VDeh was financially supported by the Belgian PRODEX program managed by the European Space Agency in collaboration with the Belgian Federal Science Policy Office.

*Author Contributions.* CG wrote the atmosphere, outgassing and escape codes, and designed the coupling between models. CG and GJG wrote the impact code. PJT wrote the StagYY code. SNR designed the N-body models and designed related simulations. CG and GJG designed the set of StagYY simulations. CG ran all simulations. All authors discussed the results and contributed to the manuscript.

**Extended data**

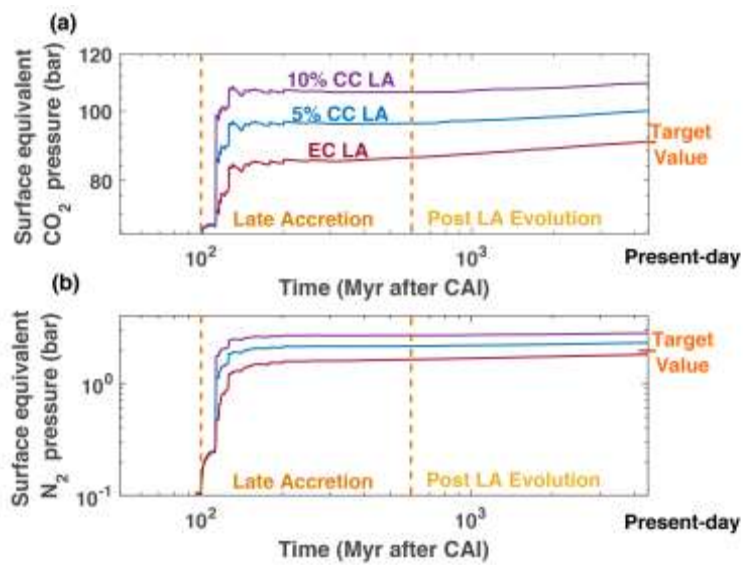

**Extended data figure 1: Evolution of CO$_2$ and N$_2$ pressure**. Time evolution of (a) CO$_2$ and (b) N$_2$ abundances in the Venus atmosphere for three different LA compositions, labelled as CC material percentage of the total LA mass delivery. MAX parameters and LA scenario D starting at 100 Myr after CAI formation are used.

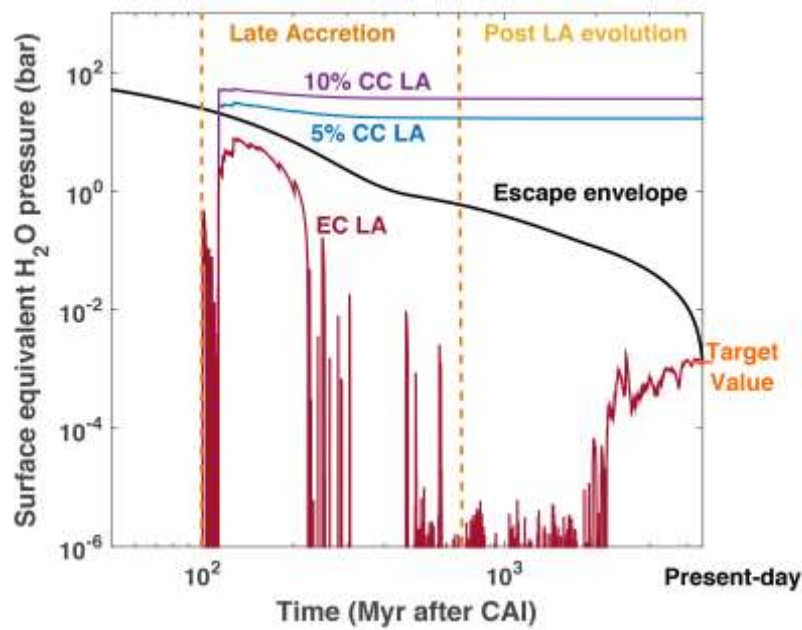

**Extended data figure 2: Evolution of water in the atmosphere of Venus.** Time evolution of H$_2$O in the Venus atmosphere for MED conditions assuming different LA compositions, labelled as CC material percentage of the total LA mass delivery. LA scenario D starting at 100 Myr after CAI formation is used.

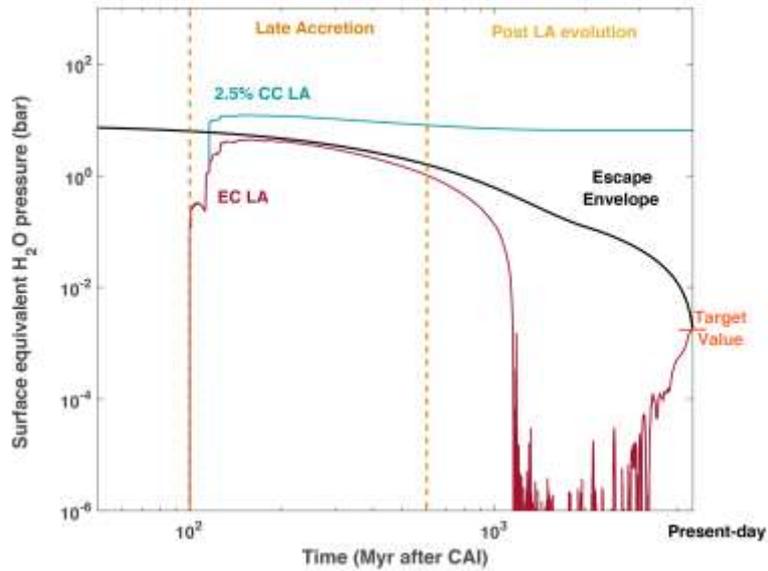

**Extended data figure 3: Evolution of water in the atmosphere of Venus.** Time evolution of $H_2O$ in the Venus atmosphere for MIN conditions assuming different LA compositions, labelled as CC material percentage of the total LA mass delivery. LA scenario D starting at 100 Myr after CAI formation is used.

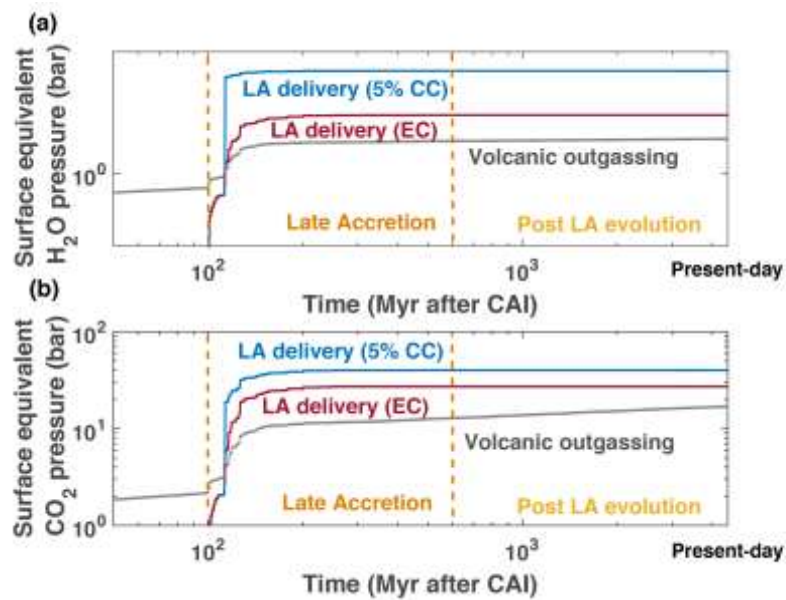

**Extended data figure 4: Comparison of delivery mechanisms.** Volcanic and impact sources for (a) $H_2O$ and (b) $CO_2$. All shown cases employ MAX parameters and LA scenario D starting at 100 Myr after CAIs.

**Extended data table 1**: List of parameters and values

| Parameter name | Parameter value |
|---|---|
| **Mantle convection modeling** | |
| Planetary radius | 6052 km |
| Planetary mass | $4.8673 \times 10^{24}$ kg |
| CMB radius | 3110 km |
| Mantle depth | 2942 km |
| Surface gravity | 8.87 m/s$^2$ |
| Surface temperature (for uncoupled cases) | 740 K |
| Initial CMB temperature | 4025 K |
| Specific heat capacity | 1200 J/kg/K |
| Latent heat of silicate melting | 600 kJ/kg |
| Reference viscosity | $10^{20}$ Pa s |
| Friction coefficient | 0.5 |
| Surface yield stress | 100 MPa |
| Internal heating at present-day | $5.2 \times 10^{-12}$ W/kg |
| Internal heating rate at model start | $18.77 \times 10^{-12}$ W/kg |
| Half-life time of radiogenic heating | 2.43 Gyr |
| **Volcanic outgassing modeling** | |
| Volatiles eruption coefficient | 10% |
| $H_2O$ concentration (initial ; present-day) | 80 ppm; 20 ppm |
| $CO_2$ concentration (initial ; present-day) | 400 ppm; 100 ppm |
| $N_2$ concentration (initial ; present-day) | 20 ppm; 5 ppm |
| Maximum eruption depth | 300 km |
| **Atmosphere and escape modeling** | |
| Solar irradiance (present-day) | 2613.9 W/m$^2$ |
| Initial $CO_2$ pressure (post magma ocean) | 65 bar |
| Initial $H_2O$ pressure (post magma ocean) | 0 bar |
| Initial $N_2$ pressure (post magma ocean) | 0 bar |
| Equilibrium temperature (present-day) | 232 K |
| Reference oxygen escape rate (non-thermal; present-day) | $6.0 \times 10^{25}$ s$^{-1}$ |
| Base of the exosphere | 200 km |
| Radius of the extended atmosphere (in planetary radii) | 8 |
| Energy deposition efficiency (hydrodynamic escape) | 15% |
| Exospheric temperature (hydrodynamic escape) | 2000 K |
| **Impact modeling** | |
| Efficiency of impact energy transfer into mantle | 0.3 |
| Impactor density | 2700 kg/m$^3$ |
| EC $H_2O$ content | 0.1% |
| EC $CO_2$ content | 0.4% |
| EC $N_2$ content | 0.02% |
| CC $H_2O$ content | 8% |
| CC $CO_2$ content | 4% |
| CC $N_2$ content | 0.2% |

**Extended data table 2**: MAX, MED and MIN specific parameter sets.

| Parameter | MAX | MED | MIN |
|---|---|---|---|
| Oxygen escape scheme (from Kulikov et al., 2006) | Case 2b | Case 2b | Case 4 |
| Multiplication factor for all mechanisms | 5 | 2.5 | 6 |
| Impact erosion scheme | Tangent plane | Tangent plane | SOVA models |